\documentclass[11pt]{article}
\newcommand{\be}{\begin{equation}}
\newcommand{\ee}{\end{equation}}

\newcommand{\ts}{\hspace{3pt}}
\newcommand{\bea}{\begin{eqnarray}}
\newcommand{\eea}{\end{eqnarray}}

\begin{document} \noindent
Reproduced from\\ FOUNDATIONS OF PHYSICS
Vol. 9, pp. 803 -- 818 (1979) \\
(with minor corrections and reformulations)

\vspace*{2.8cm}

\noindent
{\bf  QUANTUM THEORY AND TIME ASYMMETRY	}

\vskip 1.cm
\begin{quote}
\noindent
{\bf H. D. Zeh}
\vskip 0.2cm
\noindent
Institut f\"ur Theoretische Physik\\ Universit\"at Heidelberg\\ www.zeh-hd.de
\\

\end{quote}
\vskip 1.8cm

\noindent
{\bf Abstract:}
The relation between quantum measurement  and thermodynamically irreversible
processes is investigated. The reduction of the state vector is fundamentally
asymmetric in time and shows an observer-relatedness which may 
explain the double
interpretation of the state vector as a representation of physical 
states as well as
of information about them. The concept of relevance being used in all
statistical theories of irreversible thermodynamics is shown to be based on
the same observer-relatedness. Quantum theories of irreversible 
processes implicitly
use an objectivized process of state vector reduction. The conditions for the
reduction are discussed, and I speculate that the final (subjective) observer
system might even be carried by a spacetime point.
\eject

\noindent
{\bf 1. INTRODUCTION}  \vskip.5cm

\noindent Several
contributions concerned with the problem of  measurement in quantum theory
suggested a close relationship between the measurement process and irreversible
processes in statistical physics [1-4]. In fact, there are very 
plausible arguments
supporting this idea: quantum measurements seem to require irreversible
thermodynamics for the amplification of microscopic phenomena,
while the consistency problems between deterministic\footnote{The term
``deterministic" is here used to characterize dynamical laws possessing unique
solutions, although some of the founders of quantum theory 
interpreted this term (and
correspondingly its apparent opposite: the quantum mechanical 
``indeterminacy") as the
possibility of (in principle) completely determining the required 
initial conditions.}
equations of motion and master equations on the one hand, and between the
Schr\"odinger equation and the statistical character of quantum 
measurement on the
other, appear analogous in some respect. Moreover, the ensemble (incomplete
information) concept of statistical physics and the popular 
interpretation of the
wave function as  representing {\it information} about physical systems both
contain
  a similar  observer-relatedness. (In contrast, the time asymmetry 
connected with
CP violation is described by a unitary time evolution, and does {\it 
not} show any
relation to the two other time-asymmetric phenomena.)

Although the formal parts of
the above-cited investigations are essentially clear, there are deep 
differences
concerning their interpretation. There are claims that these 
contributions serve to
{\it derive} the statistical nature of quantum measurements from 
thermodynamical
fluctuations, while in general the statistical
nature of quantum theory is regarded as specific and fundamental.

Quite obviously, these discrepancies are based on different
positions concerning the basic concepts of quantum theory. It is 
generally accepted
that this theory makes statistical predictions: it describes ensembles of final
states appearing in a measurement process. This statement remains 
incomplete as long
as the conceptual terms to describe the members of these ensembles are not
defined. That is, the kinematical concepts (particle positions, wave 
functions, or
something else) to be used to describe the physical state of the 
apparatus after a
measurement (its "pointer position" or measurement result) have to be 
chosen, and the
same choice is required for the time before and during measurements in order to
formulate a dynamical theory of measurement. A specific question is 
whether classical
concepts are fundamentally required to describe the measurement 
results, or whether
they are no more than a short-hand descriptions for certain quantum 
mechanical state
vectors (``derived classical concepts"). For example, the spot resulting on a
photographic plate in the measurement of a photon position can  be described
by a stable, localized change in the corresponding molecular wave 
functions, while the
``classical" position of a pointer may be replaced by the 
sufficiently localized
center-of-mass part of the many-particle wave function for the pointer. The
probability interpretation of quantum theory then has to be expressed 
by means of a
collapse (or reduction) of the state vector.

In order to avoid conceptual confusion, some
basic interpretational issues of quantum theory are recalled in 
Section 2, before one
of them is chosen for the further discussion. In Section 3, some 
fundamentals of
statistical physics are discussed, and the role of observer-related 
concepts, such as
relevance and information, is pointed out. Their relation to the 
quantum measurement
process is studied in Section 4. Speculations about a possible solution of the
measurement problem are presented in Section 5.  \vskip.5cm

\noindent
{\bf 2. CONCEPTUAL AND DYNAMICAL DUALISM}  \vskip.5cm

\noindent
According to Niels Bohr's epistemology, quantum mechanics is
incompatible with the existence of a {\it real} microscopic world. It 
is meaningless
to ask   whether an electron ``really" is  a particle or a wave -- each choice
would lead to false conclusions. Instead, the wave function is 
assumed to be a tool
for calculating  probabilities for potential results of measurements (``pointer
positions" of a macroscopic device). These have to be described by classical
concepts,  which, however, are subject to
the uncertainty relations. During a measurement, the
electron may ``assume" a definite position or momentum (that is, a classical
property). On the other hand, quantum mechanics is assumed to be 
universally valid.
This would mean that the apparatus itself must possess a quantum 
mechanical state
vector (non-relativistically represented by a many-particle wave function).

It seems that by ``reality" (to be rejected for the microscopic world) Bohr
meant the existence of general and consistently applicable concepts to describe
states of physical objects. His ``conceptual dualism" between 
classical and quantum
concepts goes beyond the dualism characterized by conjugate variables,
such as position and momentum, or particle number and field.

According to another interpretation -- I shall
call it von Neumann's -- the state vector is regarded as
generally and exclusively applicable if in addition to
the Schr\"odinger equation another dynamical law (its ``collapse" or 
``reduction") is
assumed to apply. This collapse describes an unpredictable transition 
of a general
state vector into one of its components of a certain representation,
\be
\sum_l c_l \psi_l \to \psi_l \quad.
\ee
This existence
of two different dynamical laws may be called a ``dynamical dualism". 
The reduction is
indeterministic: the state vector allows us to predict only an ensemble of
{\it potential} state vectors at later times. In this way, the state 
vector itself
represents an objective physical state, but it contains only 
incomplete information
about the future (or past!) state. There are only intuitive rules to decide
which one of the two dynamical laws to use in a certain situation. By 
definition, the
reduction applies in ``measurement-like" situations, where the final states
are eigenstates of phenomenologically chosen observables.\footnote {So-called
measurements of the second kind, for which the collapse component 
$\psi_l$ does not
correspond to the eigenstate of the measured quantity, can be reduced 
to measurements
of the first kind by enlarging the considered system appropriately 
(for example, to
include the measurement apparatus): the ``pointer" always remains in 
its observed
state directly after being read.}

The reduction (that is, the dynamical
dualism) has to be {\it used} also in Bohr's interpretation, in particular if a
measurement serves to prepare initial states for a second one. Intuitive rules
are then again required to switch between classical variables
  and quantum mechanical state vectors.

Both interpretations appear useful pragmatically.  Von Neumann's
interpretation may be preferable for fundamental discussions, since 
it is at least
conceptually unambiguous. It does not presume any  classical concepts,
except for constructing the specific state space of a quantum
system whose classical description is known beforehand. For example, 
an individual
droplet in a Wilson chamber would not be described simply by its classical
position, but instead by a localized wave packet for the center of mass of an
aggregate of molecules. Moreover, it seems to be possible to decide on purely
dynamical grounds [5] which properties behave classically under specified
circumstances. For example, particle aspects are preferred for 
charged fermions under
``normal" conditions, but wave aspects for most electromagnetic 
phenomena or for
other bosons under conditions of superfluidity. In general, the 
quantum state of a
certain space region (which may include the apparatus and/or the observer) may
according to quantum field theory not only be found in classical field
configurations, but also in superpositions thereof.

A state vector (such as represented by a many-particle wave function) 
is usually
rejected as describing an individual physical state, since it is a 
nonlocal concept
[6]. This nonlocality of quantum theory has well known consequences 
[7,8], and it is a
direct consequence of the superposition principle. It is compatible 
with the {\it
dynamical} locality  presumed in field theories or used in non-relativistic
many-particle wave mechanics with contact interactions. The {\it kinematical}
nonlocality is at the heart of the EPR problem [9,10], while its 
consequences have
recently been experimentally verified over macroscopic distances [11] 
by using Bell's
inequality [12]. Bell's analysis has clearly revealed that {\it any} realistic
theory (for example in the sense of Einstein, Podolsky, and Rosen [9]) which is
experimentally equivalent to quantum theory must necessarily be 
nonlocal. This could
only be avoided if the hidden variables behaved in a 
thermodynamically unusual way
(conspiratorially or teleologically, that is, according to special 
final conditions).
Disregarding this latter possibility, there is thus no good reason to 
reject the
state vector as representing reality.

Since the state vector is a nonlocal concept, it cannot
generally be applied to local systems. (A local system may be defined by
means of a time-like world tube in four-dimensional spacetime.) For 
dynamical reasons,
no state vector can ever be consistently applied to a macroscopic 
system, except for
the universe as a whole [13,14], and only under special circumstances 
does it even
apply to a microscopic system. Hence, the reduction cannot be ascribed to
perturbations caused by the observer or the environment, since the 
state vector used
in Eq.\ts (1) must be assumed to represent the whole universe, thus 
already including
all possible perturbations.

The dynamical dualism may be avoided in the interpretation of quantum theory
proposed by Everett [15]. The existence of many unobserved world components --
postulated in this interpretation -- is usually regarded as an unnecessary and
extravagant complication. The assumption of ``other" components (which would
disappear according to the reduction) is in fact as unnecessary, but 
also as natural,
as the assumption of the existence of objects while not being 
observed: it follows
from an extrapolation of the empirical laws of nature (in Everett's case the
Schr\"odinger equation). However, the description of our observed universe in
derived classical terms corresponds to the {\it reduced} states, 
wherein classical
properties appear as wave packets that are approximate eigenstates of the
phenomenological observables. Therefore, von Neumann's interpretation will
be preferred to Everett's in what follows, except where explicitly 
stated. As will be
  discussed in Section 5, Everett's ``branching" is based on a similar
observer-relatedness as von Neumann's reduction.

The problem of quantum
measurement thus concerns the dynamics of the
state vector during a measurement  (the reduction).
Bohr's interpretation of the wave function would not even offer  concepts for a
non-phenomenological description of measurements. It refers to an 
outside observer --
in conflict with quantum nonlocality [14]. The ensemble of potential 
collapsed state
vectors (formally represented by a density matrix lacking the initial 
interference
terms) can evolve from the original state vector only by means of an 
indeterministic
law. Just one member of this ensemble -- the observed one -- may then 
be considered as
``real" (although this is a matter of definition, as can be seen from Everett's
interpretation). In contrast, the ensemble describing all potential results
characterizes the incomplete predictability according to this 
indeterminism. The
remaining questions then are: (a) When precisely does the reduction 
apply instead of
the Schr\"odinger equation? (b) Which interference terms disappear in a certain
situation, i.e., into which components does the total state vector 
collapse (or,
equivalently, how to justify the phenomenological observable)?

\vskip.5cm\noindent {\bf 3. THE CONCEPT
OF RELEVANCE\\ IN STATISTICAL PHYSICS}   \vskip.5cm

\noindent Statistical theories of irreversible
thermodynamics are fundamentally based on a concept of ``relevance" 
(generalized
coarse-graining) and certain initial (rather than final) conditions. 
The latter may
be  special for the relevant variables, but random with respect to the
irrelevant ones.  Statistical methods may then be used to derive 
master equations
which approximately describe the dynamics of the relevant quantities. 
The latter are
often regarded as macroscopic, and will be shown to be related to the classical
concepts in quantum theory. Master equations are able to describe the 
arrow of time
experienced by us as causality and an apparently ``fixed past". Since 
this is usually
regarded as the reason {\it why} initial conditions may be ``given", 
this argument may
appear circular from a fundamental point of view.

Master equations are based on various appropriately chosen definitions of
relevance. Examples of quantities considered as irrelevant are 
particle correlations
in Boltzmann's H-theorem, fine-grained phase-space positions in Gibbs' ink drop
analogy, or certain phase relations in quantum theories of 
irreversible processes.
Zwanzig [16] formalized the general concept of relevance by means of projection
operators $P$ in the space of phase-space densities $\rho(p, q)$ (in classical
mechanics) or density matrices $\rho_{mn}$ (in quantum mechanics). He derived a
general dynamical description (pre-master equation) of the projected densities
$\rho_{rel} = P\rho$ by methods which had been developed for special 
projections by
van Hove [17] and Prigogine [18], and he demonstrated that most 
examples studied by
the great pioneers of statistical physics could be formulated in terms of the
projection method. It was conceptually only of secondary importance 
that Zwanzig's
procedure turned out to be too general on the one hand (since not all 
projections of
density matrices are density matrices again) and too limited on the 
other (as many
important examples of relevance have to be described by
nonlinear idempotent operators [19,20]. In the following, the term ``Zwanzig
projection" is used for all appropriate idempotent operations.

The concept
of relevance introduces an observer-related element. This appears 
incompatible with
our general understanding that the laws of thermodynamics are objective. The
Zwanzig projection is able to map pure (``real") states into
(``representative") ensembles. The latter seem to correspond to the 
observer-related
concept of incomplete information, where entropy appears as a measure of lacking
information. In contrast to information theory, some objectivization 
is obtained in
statistical physics by calculating the entropy not from the actual 
information, but
by assuming that the relevant (or ``easily accessible") quantities 
are always known,
while the irrelevant ones are never known and equally distributed 
with respect to a
certain measure. For example, long-range order parameters always appear easily
accessible; this may explain the relation between entropy and 
disorder. Nevertheless,
the concept of relevance is fundamentally subjective. From an 
objective point of view
there is no reason why the position of an individual molecule should be
irrelevant, and even for practical applications may the value of 
entropy depend upon
whether one decides to consider fluctuations or not.  Although relevance also
depends on objective properties -- a quantity must be regarded as relevant if
it can easily influence other  relevant quantities --, there is a 
conceptual chain of
relevant quantities that can only end with the (potential) observer. It must be
expected that at the end of this chain (in the observer's brain) microscopic
properties are relevant again. Objective criteria by themselves, such as
dynamical stability, are certainly not sufficient, as is demonstrated 
by the general
(even classically nonlocal) constants of motion, which are quite 
irrelevant in the
sense of statistical mechanics.

Because of the
dynamical coupling between relevant and irrelevant quantities, described by
Zwanzig's reversible and still exact pre-master equation
\bea
i\rho_{rel}(t) =&& PL\rho_{rel}(t)   +  PL\,
{\rm e}\,^{[-{\rm i}(1 - P)Lt]} \rho_{irr}(0) \nonumber  \\ &&- {\rm 
i} \int d\tau
PL\, {\rm e}^{[-{\rm i} (1 - P)L\tau]} (1 - P)L\rho_{rel}(t - \tau)  \quad ,
\eea
where L is the Liouville operator, $L\rho = [H, \rho]$, entropy as a 
measure of the
(by definition missing) irrelevant information need not be constant 
in time under
deterministic equations of motion.  Total entropy may be defined 
as\footnote{See
Ref.\ts 21 for formal properties resulting for various definitions of entropy.}
\be
S :=
-k\ {\rm tr}(\rho_{rel}\ln \rho_{rel}) = -k \int d\alpha\, w_\alpha 
\ln w_\alpha +
\int da\, w_\alpha S_\alpha \quad ,
\ee
where $\alpha$ denotes the values of the
relevant quantities, while  $w_\alpha = tr(P_\alpha \rho)$, and 
$P_\alpha$ is the
projector onto the subspace corresponding to $\alpha$. The first term 
on the RHS
describes the entropy of any lacking {\it relevant} information, 
while the second one
averages accordingly over the ``physical" (objectivized) entropies 
$S_\alpha := - k
{\rm tr}(\rho_\alpha \ln
\rho_\alpha)$ as a function of $\alpha$, where $\rho_\alpha := 
P_\alpha /{\rm tr}
P_\alpha$ is independent of the ``subjective"
$\rho$. However, there is no reason for a monotonic increase of 
entropy without an
appropriate assumption about initial conditions. Thereafter, it is a 
very plausible
consequence of the large number of degrees of freedom (although hard 
to prove in
general, because of the existence of singular counterexamples) that 
the entropy of
closed systems will in practice never decrease any more. Furthermore, 
in many cases
the feedback from irrelevant information into relevant one may 
completely or partially
be neglected because of the large information capacity defined by the 
irrelevant
degrees of freedom (related to their enormous Poincar\'e cycles). This
leads to the justification of general master equations in the form
\be
\dot\rho_{rel}(t) = -PL \int_0^\infty d\tau\, {\rm e}^{[-i(1 - P)L\tau]} (1 -
P)L\rho_{rel}(t)   \quad .
\ee
  They  describe the effective
dynamics of the relevant degrees of freedom for states in accordance with the
special initial conditions. Master equations are equivalent to an alternating
application (in the ``forward" direction of time) of deterministic dynamics and
Zwanzig projections [22], whereby the latter are responsible for the 
increase of
entropy.

The
remaining {\it fundamental} problem is to understand the origin of 
the special initial
conditions. By applying statistical arguments to them, one would 
expect the relevant
quantities {\it always} -- even initially -- to be close to their 
equilibrium values.
The special conditions cannot simply be explained as being due to the 
preparation,
because the process of preparation is just an example of interacting physical
systems; the preparator himself is prepared by his environment. The initial
conditions thus have to be considered as of cosmological origin. This 
demonstrates
how the concepts of relevance are extrapolated beyond the subject's 
existence after
having been objectivized. It is certainly a nontrivial fact that the 
universe is
special in its {\it initial} conditions, and therefore asymmetric in 
time, and it
appears  difficult to justify a concept of relevance in purely objective terms.
However, it would be {\it most} remarkable if the world were 
asymmetric in terms of
precisely those variables that are relevant to an observer who later 
evolves as an
{\it effect} of these special initial conditions.

Zwanzig's method of describing the dynamics of relevant properties under
the assumption of special initial conditions seems to be completely 
analogous for
classical and quantum physics. However, this is only true for the formalism --
not for its interpretation.

The classical probability density $\rho(p,q)$ uniquely describes an
ensemble of states (points in phase space). One assumes, in this 
case, that {\it one}
of these points represents reality. The observer may then ``pick out" 
a subensemble
by a  non-disturbing observation (increase of his information). All previous
observations (his own and others'), which must also have reduced the 
initial ensemble,
have to be compatible with this final subensemble. Physically, the 
observation means
that the observer interacts with the system in such a way that some 
of his variables
change in dependence of the system's variables which distinguish between these
subensembles. The ``sensitivity" of the observer to these variables finally
determines what is relevant. This observation process should be discussed in
microscopic terms. In order to avoid a subsequent reversal of this
observation, an in practice irreversible process of information storage (in the
sense of very long Poincar\'e cycles) must accompany the observation. In
phenomenological terms one should expect the observer system to serve 
as an entropy
sink [23] in order to compensate for the entropy decrease corresponding to the
reduction of the ensemble.\footnote{BrilIouin [24]  argued that this 
entropy decrease
is compensated for by an entropy increase in the communication medium 
(for example,
light). This is in general true, but one can in principle think of a direct
interaction between system and observer.} Classically, there is no 
limit for the
capacity of such a sink, since the phase space of continuous observer
variables could be arbitrarily fine-grained, and their initial entropy has no
lower bound.

This actual information gain by an observer has to be
distinguished from a process which seems to describe a decrease in 
{\it objectivized}
entropy. To illustrate: if droplets condense out of an undercooled 
gas, their shape
and position may be regarded as macroscopic (choice of a relevance
concept), and therefore as ``given" -- regardless of an actual observation
process. The condensation into {\it definite} droplets thus seems to describe a
decrease of physical entropy, equivalent to the measure of 
information describing the
positions and shapes. However, if the transition from one droplet
position into another one -- except for their collective continuous 
motion -- is
possible only through evaporation and re-condensation, condensation 
would only then
form an irreversible process if  the number of micro-states {\it for 
each droplet
position} is larger than the corresponding number of micro-states for 
the uncondensed
gas. The entropy increase during the condensation process into the 
{\it ensemble} of
unknown droplet positions must therefore at least compensate the increase of
information during a subsequent observation of the droplet position. 
A similarly
rigorous argument cannot be found for Szillard's process of an 
actual-information
gain, although accompanying processes will always by far overcompensate it.

\vskip.5cm \noindent {\bf 4. STATE VECTOR REDUCTION\\ IN QUANTUM 
STATISTICAL PHYSICS}
\vskip.5cm

\noindent
In contrast to the
classical probability density $\rho(p,q)$, the quantum mechanical 
density matrix
$\rho_{mn}$ does not uniquely correspond to an ensemble of states. 
Although it is
often formally {\it represented} by the ensemble of orthogonal states 
$\phi_i$ which
diagonalize it (in fact, the entropy is calculated from this 
ensemble), the density
matrix defines nonvanishing probabilities also for {\it components} 
of these states.
This is true because of the fundamental probability postulate of 
quantum theory,
which states that a state $\chi$ may be found in another state $\phi$ 
(if $\langle
\phi | \chi \rangle
\neq 0$) {\it in a measurement}. Only because of this postulate can different
ensembles of not necessarily mutually orthogonal states be equivalent and
be represented by one and the same density matrix [25].

For these reasons, the process of
observation can quantum-mechanically not be described in perfect analogy
to the classical process of ``picking out". It is true that an 
element of a given
ensemble of wave functions $\{ \phi_i\}$ could be picked out by means of an
interaction of
$\phi_i
\Phi
\to
\phi_i \Phi^{(i)}$, where $\Phi$ describes the observer.  The 
observer state would
then change in dependence on the state of the observed system, and the observer
becomes aware of the property $i$. However, there are also
observations of a superposition
$\sum c_i
\phi_i$ by means of the same interaction. They are described by the 
dynamical process
\be
\sum_i c_i \phi_i \Phi_0 \to \sum_i c_i \phi_i \Phi^{(i)} \to \phi_i 
\Phi^{(i)} \quad
.
\ee
This process contains a  reduction of the wave function as the second step. The
reduction changes the state vector -- it does not merely describe the 
pick-out of some
pre-existing subensemble -- yet it cannot be described by the 
Schr\"odinger equation
[26]. This conclusion holds no matter how complicated the system $\Phi$, which
contains the observer, may be.\footnote{Since a macroscopic observer 
can never  be
kinematically isolated in quantum theory [14], he has to be 
considered as an open
system. Pragmatic theories of open systems are phenomenological and approximate
descriptions of their interactions with the environment, similar to 
master equations.
In order to write down a non-phenomenological theory, one has to make 
the ``weak
quantum cosmological assumption" that there is a state vector for the 
universe such
that
$\Phi$ in (5) describes the ``rest of the universe", including the observer.}

If one attempts, in analogy to the classical case, to introduce
an objectivized concept of relevance, and to {\it define} the entropy 
as a function of
the relevant quantities, one presumes that the latter are always given for the
real physical state. That is, one assumes that {\it the corresponding 
reductions
always occur} (without any measurement). Since this assumption 
excludes conjugate
measurements, it is equivalent to the introduction of superselection 
rules. Quantum
mechanical master equations based on a relevance concept (a Zwanzig 
projection) --
for example, by introducing a restricted set of observabIes corresponding to a
reducible algebra [27] -- therefore presuppose the reduction process. This
  reduction is thus responsible for their nonunitarity. It is then not
surprising that the master equation can be used (in a vicious circle) to
``derive" the reduction, thus erroneously indicating that an irreversible
amplification of fluctuations in the classical sense is the true cause of the
indeterminacy of quantum measurements. (The microscopic degrees of 
freedom of the
apparatus can {\it not} be the hidden variables which would determine the
measurement outcome.)

This objectivization of the reduction beyond measurements proper 
illustrates the
possibility (to be discussed further in Section 5) to {\it assume} 
that the reduction
in a measurement occurs as soon as the measured superposition has 
been amplified to
the macroscopic scale. This seems to be the reason why the objective 
existence of
classical properties has become part of our intuition. Conversely, this
impossibility of in practice distinguishing the superposition from an 
ensemble after
a measurement leads to the consequence that the reduction (or an
Everett branching) can be confirmed only by the final subjective 
observation -- in
close relationship to the application of chains of relevance ending 
at the observer in
classical statistical physics. The relation between a reduction  and
subjective awareness -- suspicious to most physicists as investigators of an
objective reality -- appears to some [28,29] as the most consistent 
interpretation,
since (a) deviations from the Schr\"odinger equation never had to  be 
used except
for measurement-like processes, and (b) fundamentally new laws may
be expected in connection with fundamentally new concepts, such as 
awareness, which
transcends physics but is obviously coupled to physics by physical processes of
observation.

Our conventional concept
of an evolving state of the universe in terms of {\it derived} 
classical terms makes
permanent use of the reduction as an indeterministic and
symmetry-violating process. The initial state of the universe may have been
completely symmetric; the reduction -- no matter when and where precisely it
occurs -- would {\it create} the complexity of the world by its
symmetry-breaking power [30].\footnote{The  expectation expressed in 
Ref.\ts 30 that
superpositions of different classical vacua may not exist does not seem to be
generally justified.} It forces ``relevant" properties to assume 
definite values by
projecting the state vector onto their corresponding subspaces. 
Although precise
rules for the reduction have never been given, these rules should 
confine and help to
define the physical meaning of relevance. The essential lesson of 
quantum theory is
that the conventional (classical) physical reality cannot be assumed to be
independent of the fundamental concepts of relevance and reduction.

\vskip.5cm \noindent {\bf 5. THE PHYSICAL EVENT OF OBSERVATION}\vskip.5cm

\noindent
A phenomenon is ``observed"
when an observer becomes aware of it. This requires the observed system to
  affect the ultimate observer system, which is known to
be localized in the brain and probably in the cerebral cortex. This 
description is
utterly nontrivial in quantum theory, since a {\it state of this 
system} cannot even
exist because of quantum nonlocality. The state of awareness or 
consciousness can
therefore not simply correspond (in the sense of a {\it naive} psycho-physical
parallelism) to ``the" state of the observer system. How can local awareness be
related to drastically nonlocal physical concepts?

The state of the final observer system (or at least its essential variables) is
{\it relevant} in an absolute (though subjective) sense. {\it 
Objectivized} concepts
of relevance and observation (for example, by an apparatus or the 
human sensorium) are
derived and of secondary nature for this purpose.

Although a quantum mechanical state vector is
nonlocal, that is,  not in general defined in terms of any state vectors of its
subsystems, quantum theory is special among nonlocal theories.  {\it Potential}
state vectors for subsystems {\it are} defined, while the general 
state of the total
system can be expanded in terms of direct products of them as a 
consequence of the
superposition principle.

Although the observer system is a subsystem of the
universe,  its relevant properties must be ``given" to the subjective observer
whenever he is aware of them. In quantum mechanical terms, this may 
be achieved by a
reduction of the state vector. Definite relevant properties of the 
ultimate observer
system are  a minimum requirement for the mechanism of the reduction 
(if there really
is one to explain definite observations). While one may {\it assume} 
that the derived
and objectivized (classical) relevant properties, too, assume 
definite values by
means of the reduction [31,32], there can be {\it no} reduction, in 
general, for
microscopic properties (such as electron positions). The borderline 
between quantum
and classical description had repeatedly to be shifted toward the 
observer -- far
beyond the microscopic realm -- whenever new experimental techniques 
allowed the
observation of  quantum mechanical phase relations. This is well 
demonstrated by
superconductivity and other long-range phenomena of quantum coherence, but most
rigorously by the quantum nonlocality experiments [11].

These considerations tend to show that in quantum theory the reduction (true or
apparent) is responsible for the fact that relevant (and therefore also
classical) properties may be ``given". Reductions
may irreversibly {\it create} initial
conditions -- they don't merely select them [25,30].
Anthropocentric reasons may be required in addition [33]  to explain 
the special
  conditions describing ``our" universe, that is, our special world
branch that has resulted from all those reductions which must have 
occurred in the
past.

Is it possible to further
constrain the mechanism of the reduction? A speculative proposal will now be
presented. If there does exist a psycho-physical coupling, it should 
be formulated
in most fundamental physical terms. Therefore, it should use strictly
quantum mechanical concepts and  remain compatible with the principles  of
relativity. This novel parallelism
(based on reduced state vectors or, alternatively, Everett's 
``relative states")
seems to form the main difference between classical and quantum theories.

The physical counterpart of
consciousness appears to be local. Let this local system be 
represented in terms of an
orthonormal basis of states ${\phi_i}$. If ${\psi_k}$ is a basis of
states for the rest of the universe, a general global state can be written as
$\sum_{ik}\phi_i
\psi_k$. According to the minimum requirement for the reduction 
mentioned above, the
reduced state after observing a result $l$, say, would be a product 
state, $\hat
\phi_l
\hat
\psi_l$, if $\phi$ describes {\it only} the ultimately relevant 
degrees of freedom.
Quite generally, the reduction is asymmetric in time, since it 
(indeterministically)
transforms an entangled state at time
$t -
\epsilon
$ into a product state at
$t +
\epsilon$. If entropy is defined to be additive (an extensive 
quantity) by choosing a
{\it local} concept of objectivized relevance, the reduction lowers this
entropy. A general theory for the mechanism of the reduction would 
have to explain or
define (a) the observer system, (b) the basis of states $\hat
\phi_l$, and (c) the time for the occurrence of a reduction (if any).

We know
that the space of states possesses an inner product, and that the 
different states
$\hat \psi_l$, observed in a certain measurement, are mutually 
orthogonal with respect
to this inner product. The factor states can then be defined 
unambiguously (except for
degeneracy) [34,35] by the plausible requirement that also the 
corresponding observer
states $\hat \phi_l$ are orthogonal, so that the total state can be 
written as the
single sum,
\be
\sum_{ik}c_{ik}\phi_i \psi_k = \sum_l \hat c_l \hat \phi_l \hat \psi_l \quad .
\ee
The transition from  the state vector (6)
to the ensemble of product states $\hat \phi_l \hat \psi_l$ with probabilities
$|c_l|^2$ would then be described by the Zwanzig projection 
corresponding to the
neglect of quantum correlations (entanglement) between the two subsystems. This
hypothesis would answer question (b), provided the observer system 
(question (a))
were given.

We also know empirically that the observer system is spatially bounded
(although we cannot give its precise boundaries), and that consciousness
changes with time.  If consciousness is in fact defined (and 
different) at every
moment of time, it should also be related to points in space: the 
truly subjective
observer system should be related to spacetime points [36]. This 
hypothesis may be
supported by the ``holographic picture" of the brain, or by Sperry's 
split-brain
experiments [37]. One would not even be in conflict with empirical 
evidence when
assuming that {\it every} spacetime point carries consciousness: we can only
communicate with some of them, and with other brains only as a whole, 
in a nontrivial
manner. The identity in time of the subjective ``I" appears as no more than a
pragmatic concept, resulting from strong causal relationships. Only 
the subjective
``I-and-Now" is required as a fundamental concept.

Before
investigating the physical consequences of this hypothesis further, 
briefly consider
question (c). This and some of the following discussion is not 
specifically based
on the radical hypothesis of point-like observer systems.

There is no natural time
interval between reductions. This difficulty is hard to overcome. Therefore,
it appears fortunate that Everett's interpretation, which does not require the
reduction, is in practice equivalent to von Neumann's. One may then apply
Erhard Schmidt's canonical representation (6) to the never reduced 
state vector at
every moment of time in order to postulate [14] that consciousness is 
``parallel" to
one of the states $\hat \phi_l$  (or to all of them separately).\footnote{The
branching with respect to different ``memory states", proposed by 
Everett, would
correspond to an objectivized reduction.} This new hypothesis avoids 
the dynamical
time asymmetry of the reduction. As discussed in a previous paper 
[38], Everett's
model (which corresponds to a pre-master equation) is 
indistinguishable in practice
from von Neumann's interpretation
only if {\it all} unobserved (nonexisting in von Neumann's 
interpretation) components
$l'
\neq l$ which might {\it later} interfere with the observed one 
possess negligible
amplitudes
$c_{l'}$. In this way, the
time-asymmetric dynamical law of reduction is replaced by an 
asymmetric assumption
about cosmological initial conditions for the total Everett wave 
function [36]. The
space of unobserved Everett branches has to be sufficiently empty in 
order to serve
as a perfect sink for initially local quantum phases. (One may have 
to use quantum
gravity in order to decide if this emptiness is related to
  the black night sky or the thermodynamical
arrow of time [39,40].)

Consider a space-like hypersurface of spacetime described by a
constant time coordinate $t$, and assume that the general state vector
$|\alpha(t)\rangle$ on this hypersurface can be written as a wave functional
$\Psi[\phi_\nu({\bf x} ),t]$,
\be
|\alpha(t)\rangle = \int {\mathcal D} \phi_\nu ({\bf x} )\, 
\Psi[\phi_\nu ({\bf x}
),t] \, |\phi_\nu ({\bf x} ) \rangle \quad ,
\ee
  where
$|\phi_\nu ({\bf x }) \rangle$ is a state with definite (classical) 
amplitudes of
certain fundamental fields $\phi_\nu ({\bf x }), \nu = 1,2,..., N$. 
If $I_1$ is some
region of space, with $ I_2$ being its complement, $|\phi_\nu (\bf x 
) \rangle$ is
approximately (neglecting Casimir-type entanglement of the 
relativistic vacuum) a
direct product,
$|\phi_\nu ({\bf x })
\rangle_{I_1}
\,|\phi_\nu ({\bf x })
\rangle_{I_2}$. The Schmidt  representation with respect to $I_1$ 
and $I_2$ reads
\be
|\alpha(t)\rangle = \sum_l \hat c_l |\hat \beta_l(t) \rangle_{I_1} 
|\hat \gamma_l(t)
\rangle_{I_2} \quad ,
\ee
where the factor states can
again be written as  superpositions of field eigenstates,   for example
\be
|\hat \beta_l(t) \rangle_{I_1} = \int_{(I_1)} {\mathcal D} \phi_\nu 
({\bf x} )\,
\hat \Psi_l[\phi_\nu ({\bf x} ),t] \, |\phi_\nu ({\bf x} ) 
\rangle_{I_1} \quad .
\ee
The functional integral $\int_{(I_1)} {\mathcal D} \phi_\nu ({\bf 
x})\dots$ now runs
over all field amplitudes on the space region
$I_1$ only. Provided $I_1$ carries precisely the ultimate observer
system, one might postulate a (partly reversible) branching of 
consciousness into
these product states which are distinguished by their index
$l$.

If the region $I_1$ shrinks to a
point ${\bf x}_0$, say, the states $| \phi_\nu ({\bf x} \rangle_{I_1}$ become
\be
| \phi_\nu ({\bf x}) \rangle_{\bf x_0} = |\xi_\nu \rangle := |\phi_\nu ({\bf
x_0})\rangle \quad ,
\ee
that is, states characterized by the values of all fields at the 
point ${\bf x_0}$.
    The quantum states of such a point can therefore be written as
\be
| \beta \rangle_{\bf x_0} = \int d\xi_1 \dots d\xi_N \, f(\{ \xi_\nu 
\}) \, | \{
\xi_\nu\}
\rangle
\ee
The manifoldness
of these superpositions appears rich enough to represent primitive 
conscious awareness
in a psycho-physical parallelism. It seems that quantum 
superpositions have never
been considered, for example, in neuronal models, since only 
classical states of
definite neuronal excitation are usually taken into account. These 
quasi-classical
states are also measured by external
neurobiologists. Quantum theory would admit their superpositions, too, thus
giving rise to a far greater variety of physical states which may be 
experienced by
the subjective observer. (Note added: When used for information {\it 
processing}, such
superpositions would now be called ``quantum bits". As demonstrated 
by M.\ts Tegmark,
they can {\it not} be relevant for neuronal and similar processes in 
the brain -- see
Phys.\ts Rev {\bf E61}, 4194 (2000) or quant-ph/9907009.)

However, the physical carrier of states of primitve consciousness can 
neither be
expected to include unconscious memories, nor those neuronal
activities which are related to ``behavior" (such as speech). 
Evidently, most brain
activities remain unconscious, but nonetheless contribute to the complexity of
``conscious behavior". For example, the quantum state on the region
$I_2$, which by definition is external to the ultimate observer 
system, may be further
divided into the rest of the brain,
$I_{21}$, and the external world in the usual sense,
$I_{22}$. The states
$|\hat  \gamma_l \rangle_{I_2}$ may then again be written as
\be
|\hat \gamma_l \rangle_{I_2} = \sum_m \hat d_m^{(l)}\, |\hat \delta_m^{(l)}
\rangle_{I_{21}} |\hat  \epsilon_m^{(l)} \rangle_{I_{22}} \quad .
\ee
The type of states $|\hat \delta_m^{(l)} \rangle_{I_{21}} $ is thus essentially
determined by the unavoidable and mostly irreversible interaction of 
the brain with
its environment
$I_{22}$. The observer system
$I_1$ -- perhaps assumed to interact only with $I_{21}$ -- cannot, 
therefore, observe
relative phases between states of different $m$ (see Section 4 of 
Ref.\ts 38). If the
states on
$I_{21}$ possess memory properties, which must be ``robust" or 
dynamically stable
[14], these properties (in contrast to the states which carry 
consciousness) would
behave classically: memory is irreversibly objectivized by means of 
the unavoidable
interaction with its environment.
\eject

\vskip.5cm \noindent {\bf REFERENCES} \vskip.5cm

\noindent 1. G. Ludwig, Z. Physik {\bf 135}, 483 (1953). \\
2. A. Danieri, A. Loinger, and G. Prosperi, Nucl. Phys. {\bf 33}, 297 
(1962). \\
3. W. Weidlich, Z. Physik {\bf 205}, 199 (1967). \\
4. I. Prigogine and L. Rosenfeld, Nature {\bf 240}, 25 (1972). \\
5. 0. K\"ubler and H. D. Zeh, Ann. Phys. ( N. Y.) {\bf 76}, 405 (1973). \\
6. W. Heisenberg, Physik und Philosophie (Hirzel, Stuttgart, 1972). \\
7. B. d'Espagnat, Conceptual Foundations of Quantum Theory (Benjamin, New York,
1971).\\
  8. D. J. Bohm and B. J. Hiley, Found. Phys. {\bf 3}, 93 (1975). \\
9. A. Einstein, N. Rosen, and B. Podolski, Phys. Rev. {\bf 47}, 777 (1935). \\
10. D. Bohm and Y. Aharonov, Phys. Rev. {\bf 108}, 1070 (1957). \\
11. J. F. Clauser and A. Shimony, Rep. Prog. Phys. {\bf 41}, 1881 (1978). \\
12. J. S. Bell, Physics {\bf 1}, 195 (1964). \\
13. K. Baumann, Z. Naturforsch. {\bf A25}, 1954 (1970); Acta Phys. 
Austr. 36, {\bf 1}
(1972).  \\
14. H. D. Zeh, Found. Phys.  {\bf 1}, 69 (1970) -- reprinted in J.A. 
Wheeler and W.H.
Zurek, Quantum Theory and Measurement (Princeton UP, 1983).
\\ 15. H. Everett III, Rev. Mod. Phys. {\bf 29}, 454 (1957). \\
16. R. Zwanzig, in Boulder Lectures in Theoretical Physics, Vol.\ts 
{\bf 3} (1960), p.
106.\\ 17. L. Van Hove, Physica {\bf 22}, 343 (1956). \\
18. I. Prigogine, Nonequilibrium Statistical Mechanics (New York, 1962). \\
19. P. N. Argyres and P. L. Kelly, Phys. Rev. {\bf 134}, A98 (1964). \\
20. R. M. Lewis, J. Math. Phys. {\bf 8}, 1448 (1967). \\
21. A. Wehrl, Rev. Mod. Phys. {\bf 50}, 221 (1978). \\
22. R. Mirman, Found. Phys. {\bf 5}, 491 (1975). \\
23. L. Szilard, Z. Physik {\bf 53}, 840 (1929). \\
24. L. Brillouin, Science and Information Theory (Academic, New York, 1956). \\
25. H. D. Zeh, in Proc. {\bf 49}th Enrico Fermi School of Physics, B. 
d'Espagnat, ed.
(Academic, New York, 1972), p. 263. \\
26. E. P. Wigner, Am. J. Phys. {\bf 31}, 6 (1963). \\
27. I. M. Jauch, Foundations of Quantum Mechanics (Addison-Wesley, 
Reading, Mass.,
1968). \\
28. F. London and E. Bauer, La theorie de I'observation mecanique quantique
(Hermann, Paris, 1939). \\
29. E. P. Wigner, in The Scientist Speculates, L. I. Good, ed. 
(Heinemann, London,
1962). \\
30. H. D. Zeh, Found. Phys. {\bf 5}, 371 (1975). \\
31. Ph. Pearle, Phys. Rev. {\bf D13}, 857 (1976); Intern. J. Theor. 
Phys. {\bf 18},
489 (1979).  \\
32. D. Bedford and D. Wang, Nuovo Cimento {\bf 26B}, 313 (1975). \\
33. C. Misner, K. S. Thorne, and I. A. Wheeler, Gravitation (Freeman, 
San Francisco,
1973), p. 1217. \\
34. E. Schmidt, Math. Annalen 63, {\bf 433} (1907). \\
35. E. Schr\"odinger, Proc. Cambr. Phil. Soc. {\bf 31}, 555 (1935). \\
36. H. D. Zeh, in Proc. of the 4th Conference on the Unity of the 
Sciences (Int.
Cultural Foundation, New York, 1975), p. 287. \\
37. R. W. Sperry, in The Neurosciences Third Study Progrum, F. 0. 
Schmidt and F. G.
Worden, edts. (MIT Press, Cambridge, Mass., 1974). \\
38. H. D. Zeh, Found. Phys. {\bf 3}, 109 (1973) -- quant-ph/0306151. \\
39. T. Gold, ed., The Nature of Time (Cornell, Ithaca, N. Y., 1967). \\
40. B. Gal-Or, Modern Developments in Thermodynamics (Wiley, New York, 1974).

\end{document}